# Angle-resolved photoemission study of cobalt oxide superconductor $Na_xCoO_2 \cdot yH_2O$ : Observation of the Fermi surface

T. Shimojima[1,*], K. Ishizaka[1], S. Tsuda[1], T. Kiss[2], T. Yokoya[3], A. Chainani[4], S. Shin[1,4], P. Badica[5,6], K. Yamada[5], K. Togano[7]

[1] *Institute for Solid State Physics (ISSP), University of Tokyo, Kashiwa, Chiba 277-8581, Japan*

[2] *The Institute of Physical and Chemical Research (RIKEN),Wako, Saitama 351-0198, Japan*

[3] *The graduate school of Natural science and Technology, Okayama University Tsushima, Okayama 700-8530, JAPAN*

[4] *The Institute of Physical and Chemical Research (RIKEN),Sayo-gun, Hyogo 679-5148, Japan*

[5] *Institute for Materials Research, Tohoku University, 2-1-1 Katahira, Aoba-ku, Sendai, 980-8577Japan*

[6] *National Institute of Material Physics, Bucharest, POB MG-7,RO-76900, Romania*

[7] *National Institute for Materials Science, 1-2-1 Sengen, Tsukuba 305-0047 Japan*

**Abstract**

**Cobalt oxide superconductor $Na_xCoO_2 \cdot yH_2O$ is studied by angle-resolved photoemission spectroscopy (ARPES). We report the Fermi surface (FS) topology and electronic structure near the Fermi level ($E_F$) of $Na_xCoO_2 \cdot yH_2O$. Our result indicates the presence of the hexagonal FS centered at $\Gamma$ point, while the small pocket FSs along $\Gamma$-K direction are absent similar to $Na_xCoO_2$. The top of the $e_g$' band, which is expected in band calculations to form the small pocket FSs, extends to within 20 - 50 meV below $E_F$, more closer to $E_F$ than in $Na_xCoO_2$. We discuss its possible role in superconductivity, comparing with other experimental and theoretical results.**

*To whom correspondence should be addressed.    Email address: t-shimo@issp.u-tokyo.ac.jp

The discovery of superconductivity in $H_2O$ intercalated transition metal oxide $Na_xCoO_2 \cdot yH_2O$[1] sheds new light on the possible exotic pairing mechanism arising in strongly correlated system. While sharing the relevance of a two-dimensional structure with high-$T_c$ cuprates and ruthenates, it has a special aspect: the triangular network of $Co^{3+}$(s = 0) and $Co^{4+}$(s = 1/2), which would give rise to a topological frustration in case the spin and charge degree of freedom are fairly localized. The interplay of the superconductivity and its crystal, electronic, and magnetic structures is still under active discussion, both theoretically and experimentally.

Regarding the non-hydrated $Na_xCoO_2$ system, the physical properties have been quite extensively investigated: it shows large thermopower (x = 0.5),[2] magnetic order (x ≥ 0.75),[3] insulating phases (x = 0.25, 0.5),[4] and so on. Such a rich phase diagram[5] directly arises from the electronic structure near $E_F$, namely the quasiparticle dispersion and the FS topology. In order to investigate them, systematic studies using ARPES has been reported.[6-9] The conduction band of $Na_xCoO_2$ mainly consists of Co $t_{2g}$ orbitals which split into $a_{1g}$-singlet and $e_g$'-doublet due to a trigonal crystal field. According to the local density approximation (LDA) band calculation,[10] they form the hexagonal FS centered at Γ point and the small-pocket-like FS along Γ-K direction, respectively. Theoretically, this $e_g$' derived small pocket FS is responsible for ferromagnetic fluctuations giving a pairing mechanism for triplet superconductivity.[11] The first ARPES study on the $Na_{0.7}CoO_2$ by Hasan *et al.*,[6] however, observed a single hexagonal $a_{1g}$ FS without the evidence of the small pocket. Very recent, systematic investigation on the $Na_xCoO_2$ (x = 0.3 ~ 0.72) also revealed the *x*-dependent $a_{1g}$ FS evolution whose volume alone satisfies the Luttinger theorem, confirming absence of the small pocket.[8] The preceding ARPES results on non-hydrated $Na_xCoO_2$ show that the top position of the $e_g$' band is x-independent, sinking 100 ~ 200 meV below $E_F$,[7-9] suggesting that its role will not appear in low energy phenomena, such as superconductivity. To investigate the background of the superconductivity, however, it is highly desirable to study the superconducting pairing mechanism from the electronic structure of the $Na_xCoO_2 \cdot yH_2O$ itself. ARPES on sodium cobalt oxides thus far has been limited to non-hydrated $Na_xCoO_2$, possibly due to the difficulty of handling hydrated crystals under ultra high vacuum (UHV) without loss of water. Moreover, until now, there has been no experimental study that could clarify the FS topology of superconducting $Na_xCoO_2 \cdot yH_2O$.

In this work, we have ensured the retention of $H_2O$ in the crystal and superconducting transition temperature $T_c$ before and after measurements, and performed the first ARPES on superconducting $Na_xCoO_2 \cdot yH_2O$ to obtain the electronic structure near $E_F$. The presence of the large hexagonal FS derived from $a_{1g}$ band is evident, which is similar to that of non-hydrated $Na_xCoO_2$. On the other hand, the $e_g$' band which is expected in band calculations to form the small pocket FSs, extends much closer up to $E_F$ than in the non-hydrated $Na_xCoO_2$. These results will give the first step to the determination of pairing symmetry in this hydrated superconductor.

Single crystal $Na_xCoO_2 \cdot yH_2O$ is synthesized from $Na_{0.7}CoO_2$ grown by floating zone method, as described in Ref. 12. The hydrated crystal loses its water molecules in ambient atmosphere and room temperature. Since ARPES measurement is performed under UHV, we have followed a careful procedure as follows: First, the single crystal of $Na_xCoO_2 \cdot yH_2O$ was transferred into the vacuum chamber with its temperature kept at ~250 K. Second, the crystal was cleaved under UHV using adhesive tape at 230 K. We have checked that water molecules in the crystal are stable against desorption below

~250 K even in the UHV. In a previous PES study on the O1s core level of $Na_xCoO_2 \cdot yH_2O$, we have observed a peak feature derived from oxygens in $H_2O$ molecules, which are chemically bonded in the crystal.[13] In the present study, the pressure did not change, throughout the cleaving procedure and the measurements. We have also confirmed $T_c$ of 4 K in the samples before and after the ARPES measurements using a Magnetic Property Measuring System (MPMS, Quantum Design). All ARPES measurements were performed on a spectrometer built using a GAMMADATA-SCIENTA SES2002 electron analyzer and a high-flux helium-discharging lamp with a toroidal grating monochromator. The total energy resolution of the He I$\alpha$ (21.2 eV) resonance line was set to 40 meV to get a reasonable count rate of photoelectrons. The sample temperature was set at 30K, which is measured using a silicon-diode sensor. The base pressure of the analyzing chamber was better than $5 \times 10^{-11}$ Torr. $E_F$ of sample was referenced to that of a gold film evaporated near the sample substrate. All measurements were carried out within 3 hours after the cleaving and we have confirmed the reproducibility of the data.

ARPES detects the energy and momentum dependence of electrons emitted from a solid by photon irradiation. The energy distribution curve (EDC) at a fixed momentum and momentum distribution curve (MDC) at a fixed energy show peak structures due to band dispersions. ARPES can directly probe the FS by measuring the MDC at $E_F$. In Fig. 1(a), momentum space measured by ARPES in this work, numbered as line 1 to 6, are shown on the FS of $Na_xCoO_2$ obtained by LDA calculations.[14] While line 6 (red) is near Γ-M line, line 2 (blue) cuts across the center of the calculated small pockets. Both lines also cross the hexagonal FS at different momentum points. Corresponding EDCs are shown in Figs. 1(b)-(d). In Fig. 1(c), raw EDCs of $Na_xCoO_2 \cdot yH_2O$ along line 2(blue line) are shown with those along line 6(broken red line). We can clearly see larger spectral intensity for the line2 (blue) in comparison with line 6 (red) from the momentum region k = 0.52 Å$^{-1}$ to k = 1.0 Å$^{-1}$, while both spectra almost coincide with each other at k = 0.0 Å$^{-1}$. In order to distinguish this difference more clearly, we subtract a background spectrum from raw EDCs. For background, we used the EDC near Γ point, which does not contain any band dispersion. After this procedure, as illustrated in Fig.1 (b), two kinds of band dispersions can be separated as shown in Fig.1 (d). At k = 0.0 Å$^{-1}$, analyzed EDCs show almost flat feature, indicating the absence of any band dispersion. As k increases, analyzed EDC changes its spectral weight systematically. Then, we plotted the peak positions estimated by fitting MDCs with a Lorentzian function, which are more reliable than EDCs. Both blue and red diamonds are superimposed on the slow dispersions of the $a_{1g}$ band, crossing $E_F$ at different momentum points for line2 and line6 respectively. In addition, light green diamonds indicate dispersive $e_g$' band on line2, which approach $E_F$ but do not cross $E_F$. To get the image of the dispersions, we analyzed the data by taking the second derivative of ARPES intensity I(k,ω) with respect to momentum k. The contour plot images of $d^2I/dk^2$ are shown in Fig.1(e). Numbers on the panels corresponds to that of the lines measured. A white arrow on each panel indicates the Fermi momentum $k_F$, where the $a_{1g}$ band crosses $E_F$. Diamond markers are used as the indicator of band dispersions in panel 2 and 6 similarly as in Fig. 1(d). Six panels show systematic changes in intensity. The intensity due to the $e_g$' band (indicated with light green diamonds in panel 2) gradually decreases and moves leftward from panel 2 to panel 6. On the other hand, the intensity reflecting the $a_{1g}$ band and $k_F$ move rightward from panel 1 to 6. Thus, we can separate the two kinds of the band dispersions in Fig. 1(e). Highest intensity region mostly indicates the cross point of the

$a_{1g}$ and the $e_g$' bands. However it is difficult to unambiguously separate the two bands near $E_F$ due to the thermal broadening and/or insufficient energy resolution.

Next we will see how these two bands are located in the Brillouin zone(BZ). In Figure 2(a)-(f), we show the MDC intensity images taken at several binding energies (from $E_F$ up to 50 meV binding energy, steps of 10meV) by integrating the spectral weight with an energy window of ± 2 meV. The MDC intensity image at $E_F$ represents the FS, which indicates the presence of hexagonal FS centered at Γ point and the absence of small pockets along Γ-K direction. At 10 meV binding energy, the region surrounded by the hexagonal $a_{1g}$ band becomes slightly larger than that at $E_F$, due to the hole-like character of the $a_{1g}$ band. We can also see the systematic evolution of the small high-intensity region along Γ-K direction from 20 meV to 50 meV binding energy. It indicates that the top position of the $e_g$' band lies near 20 – 50 meV below $E_F$. Recent ARPES study on $Na_xCoO_2$ (x = 0.3 ~ 0.72) has reported that the top of the $e_g$' band sinks about 100 ~ 200 meV below $E_F$[7-9]. While it is not clear at present why the band top position differs between $Na_xCoO_2 \cdot yH_2O$ and $Na_xCoO_2$, one possible scenario is as follows. Absence of the small pockets among $Na_xCoO_2 \cdot yH_2O$ and $Na_xCoO_2$ is explained theoretically by the strong electronic correlation effect which tends to push down the $e_g$' band.[15] On the other hand, shallower $e_g$' band in the $Na_xCoO_2 \cdot yH_2O$ can be attributed to the increase of the trigonal crystal field due to the distortion along the c-axis, which tends to push up $e_g$' bands.[11] Indeed, the structure analysis of $Na_xCoO_2 \cdot yH_2O$ indicates a thinner conducting layer consisting of the O-Co-O trilayer,[16] and a higher trigonal crystal field, than that of non-hydrate $Na_xCoO_2$. Such difference in the depth of the $e_g$' band top may be an important point regarding the paring mechanism in this system, as we discuss later.

Here we discuss the FS volume in comparison with the valence of cobalt ion, employing the Luttinger theorem. Fig. 3 shows same FS as shown in Fig. 2(a). The Luttinger theorem relates the carrier density X and FS volume in the first BZ. The Luttinger theorem in 2 dimensaional (2D) system is represented as $X=X' \equiv 1 - 2S_{FS}/S_{BZ}$, where $X'$ is the effective carrier density, $S_{FS}$ is the area surrounded by the 2D FS, and $S_{BZ}$ area of the first BZ. Recent detailed chemical study[17] by Takada *et al.* indicates the accurate chemical formula of the superconducting cobalt oxide is $Na_{0.343}(H_3O)_{0.237}CoO_2 \cdot 1.19H_2O$, considering the substitution of $Na^+$ by $H_3O^+$ during the water intercalation. Then carrier density X is estimated to be 0.58, instead of 0.35 as estimated by the previous formula of $Na_{0.35}CoO_2 \cdot 1.3H_2O$. The valences of cobalt ion for respective cases are calculated as +3.42 and +3.65. In Fig. 3, the virtual Luttinger FSs, assuming the simple hexagons, are shown by blue and yellow lines for respective cases described above. Light green markers between them indicate the actual peak positions estimated from the MDCs at $E_F$ in this work, showing the $k_F$ on Γ-K (0.68 Å$^{-1}$) and Γ-M direction (0.78 Å$^{-1}$). However, there is an analytical error of ± 0.035 Å$^{-1}$ arising from the six-fold symmetrizing procedure. Assuming that the Luttinger theorem holds in this material, $X'$ is calculated to be ~0.44 (± 0.05), which corresponds to the nominal cobalt valence of +3.56 (± 0.05). This result shows an intermediated value between those expected by the above-mentioned two chemical formulas.

Finally, we discuss several implications presented by our results. We can conclude that the hexagonal FS centered at Γ point is present and the small pockets on the Γ-K direction do not exist in superconducting $Na_xCoO_2 \cdot yH_2O$. Theoretically, some authors propose the d+id symmetry (which actually has six-fold symmetry) within RVB theory assuming only the $a_{1g}$ hexagonal FS.[18-22] In another case, Yada *et al.* propose the possibility of s-wave superconductivity enhanced by the interband($a_{1g} \leftrightarrow e_g'$) hopping of Cooper pairs caused by the coupling with optical phonons.[23] In this scenario, $e_g'$ bands do not necessarily cross $E_F$, but should be located within binding energy less than the phonon energy $\hbar\omega_{ph} \sim 70$ meV. Our result can support these scenarios, possibly producing the singlet superconducting state with s or d + id symmetries.

Nevertheless, we cannot entirely rule out the possibility that our results reflect a kind of surface effect. Incident light of 21.2eV creates photoelectrons with escape depth of ~10Å,[24] which corresponds to only 1 or 2 $CoO_2$ layers from the cleaved surface. If the surface effect exists, photoemission results will be different from the bulk electronic structure. Two types of surface effects can be considered for the cobalt oxides. One is the disorder induced on the surface cleaved at low temperature, which can be removed by thermal cycling.[8] In this work, it is not so relevant because the sample temperature ranged from 230 K to 30K after the cleaving procedure. Actually, we did not observe any double-band dispersion due to the disorder as reported in Ref. 8. The other possibility is a relaxation of the trigonal distortion on the cleaved surface. In this material, electronic structure near $E_F$, especially regarding the location of $e_g'$ bands, is reported to be sensitive to the trigonal crystal field in the $CoO_6$ octahedrons. Theoretical study indeed indicates that the increase of the trigonal crystal field pushes up $e_g'$ bands to generate the small pockets.[11] If there exists a strong lattice relaxation at the cleaved surface, it may modify the intrinsic electronic structure resulting in the absence of the small pockets in our result. Further studies using a photon source producing longer escape depth of photoelectrons are required to exclude surface effects and make a definitive conclusion.

In conclusion, we have studied the FS topology and electronic structure near $E_F$ of

$Na_xCoO_2 \cdot yH_2O$ for the first time by ARPES. Our result indicates the presence of the single hexagonal FS centered at $\Gamma$ point without the small pocket FSs along $\Gamma$-K direction similar to the non-hydrated $Na_xCoO_2$. The top position of the $e_g'$ band, which is predicted by LDA calculation to form the small pocket FSs, is located 20 - 50 meV below $E_F$. This is much more closer to $E_F$ than that observed in non-hydrated $Na_xCoO_2$, and may be an important difference regarding the pairing mechanism. Further studies with a photon source producing longer escape depth are still desirable for entirely excluding the concern on surface effects.

**Acknowledgements:** We thank M. Ogata, Y. Yanase, M. Mochizuki, H. Kontani and K. Yada for very valuable discussions.

**Figure captions**

**Figure 1** Angle-resolved photoemission results of $Na_xCoO_2 \cdot yH_2O$. (a) Fermi surfaces obtained by LDA calculations[14] and the momentum space measured by ARPES, numbered from line 1 to 6. (b) Procedure for subtracting background from raw EDCs. Broken black line is the background spectrum (EDC measured near Γ point which does not contain any band dispersion). Upper thin and lower thick spectra show raw data and analyzed spectra, respectively. (c) Raw EDCs along the line 2 (blue curve) and line 6 (broken red curve). (d) EDCs after subtracting the background as illustrated in (b). Both blue and red diamonds are superimposed on the slow dispersions of the $a_{1g}$ band for line 2 and line 6 respectively. Light green diamonds indicate dispersive $e_g$' band on line 2. All the peak positions are determined from fits to MDCs. (e) Second derivative of ARPES intensity $I(k,\omega)$ with respect to momentum k. Number on the panels corresponds to the measured regions indicated in (a). Markers in the panel 2 and 6 represent those in (d). White arrows on each panel indicate the Fermi momenta $k_F$, where the $a_{1g}$ band crosses the $E_F$.

**Figure 2** Intensity images at the momentum space ranging from $E_F$ and up to 50 meV binding energy (steps of 10meV), obtained by integrating spectral weight with an energy window of ± 2 meV. Only hexagonal FS centered at Γ point is observed at $E_F$. Systematic evolution of high-intensity region at Γ-K direction is observed from 20 meV to 50 meV binding energy, which indicates that the top position of the $e_g$' band is located within 20 - 50 meV below $E_F$.

**Figeure 3** Fermi surface of $Na_xCoO_2 \cdot yH_2O$ with virtual FSs estimated from the cobalt valencies as reported by Takada *et al.*.[1,17] Blue and yellow hexagons indicate the FS corresponding to the Co valence of +3.42 ($Na_{0.343}(H_3O)_{0.237}CoO_2 \cdot 1.19H_2O$) and +3.65

($Na_{0.35}CoO_2 \cdot 1.3H_2O$), respectively. Green markers show the peak positions obtained from MDC at $E_F$, which indicates the Fermi momenta $k_F$ of ~0.68 Å for Γ-K and of ~0.78 Å for Γ-M direction, respectively.

**Figures**

Fig.1

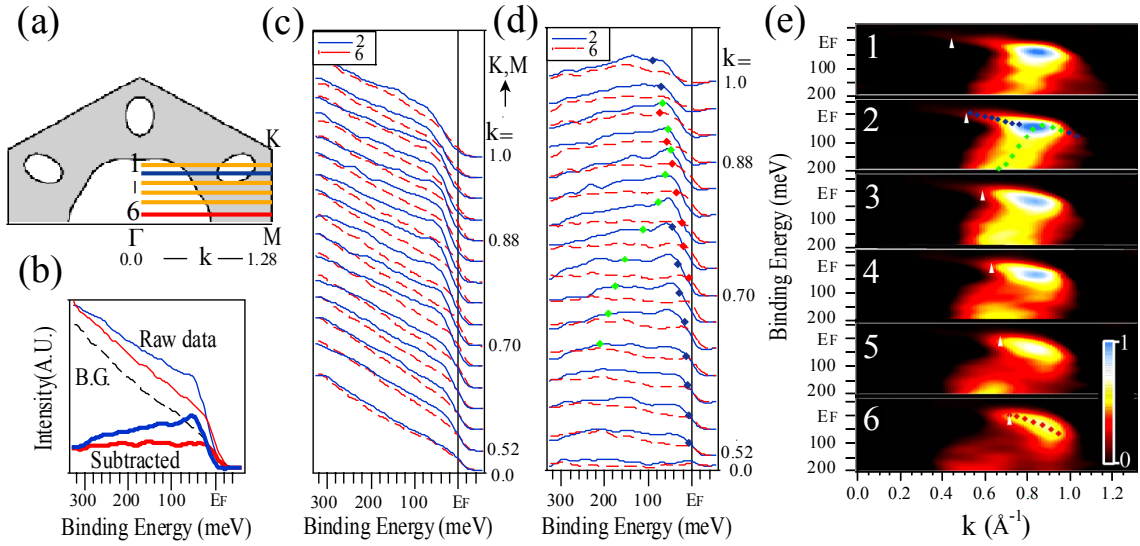

T. Shimojima et al.,

Fig.2

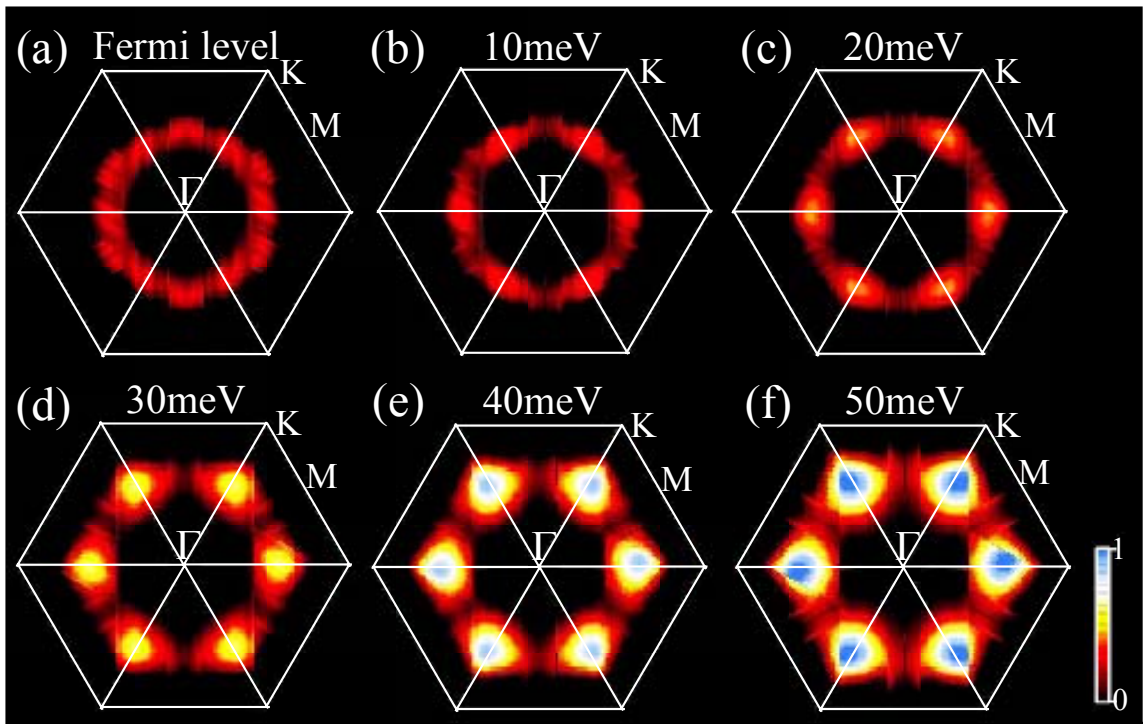

T. Shimojima et al.,

Fig.3

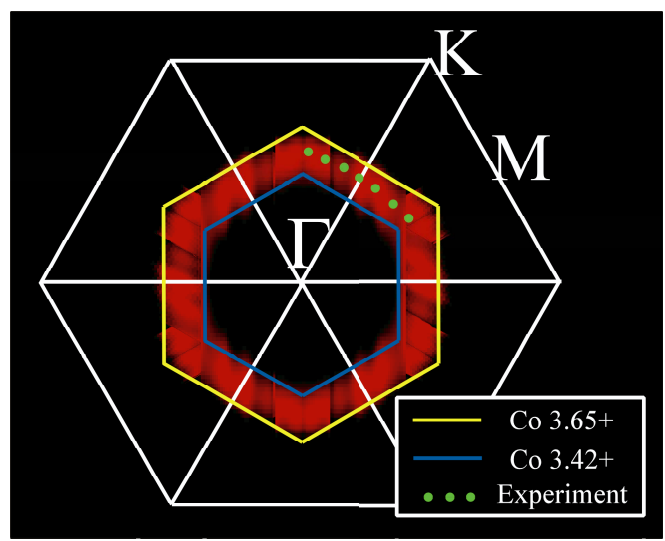

T. Shimojima et al.,